\documentclass[aps,prl,reprint,superscriptaddress,longbibliography]{revtex4-1}
\usepackage{blindtext}
\usepackage{graphicx}
\usepackage{amsmath,amssymb}
\usepackage{multirow}
\usepackage{url}
\usepackage{color,soul}
\usepackage[colorlinks,
            linkcolor=blue,
            anchorcolor=blue,
            citecolor=blue,
            ]{hyperref}
\usepackage{setspace}

\begin{document}

\title{Short-wavelength Reverberant Wave Systems for Physical Realization of Reservoir Computing}

\author{Shukai Ma}
\email{skma@umd.edu}
\affiliation{Quantum Materials Center, Department of Physics, University of Maryland, College Park, Maryland 20742, USA}
\author{Thomas M. Antonsen}
\affiliation{Department of Physics, University of Maryland, College Park, Maryland 20742, USA}
\affiliation{Department of Electrical and Computer Engineering, University of Maryland, College Park, Maryland 20742-3285, USA}
\author{Steven M. Anlage}
\affiliation{Quantum Materials Center, Department of Physics, University of Maryland, College Park, Maryland 20742, USA}
\affiliation{Department of Electrical and Computer Engineering, University of Maryland, College Park, Maryland 20742-3285, USA}
\author{Edward Ott}
\affiliation{Department of Physics, University of Maryland, College Park, Maryland 20742, USA}
\affiliation{Department of Electrical and Computer Engineering, University of Maryland, College Park, Maryland 20742-3285, USA}

\begin{abstract}

Machine learning (ML) has found widespread application over a broad range of important tasks. 
To enhance ML performance, researchers have investigated computational architectures whose physical implementations promise compactness, high-speed execution, physical robustness, and low energy cost.
Here, we experimentally demonstrate an approach that uses the high sensitivity of reverberant short wavelength waves for physical realization and enhancement of computational power of a type of ML known as reservoir computing (RC).
The potential computation power of RC systems increases with their effective size.
We here exploit the intrinsic property of {short wavelength reverberant wave} sensitivity to perturbations to expand the effective size of the RC system by means of spatial and spectral perturbations.
Working in the microwave regime, this scheme is tested {experimentally} on different ML tasks. 
Our results indicate the general applicability of reverberant wave-based implementations of RC and of our effective reservoir size expansion techniques.

\end{abstract}

\maketitle

\begin{spacing}{1}

\section{I. Introduction}

{Machine learning (ML) algorithms have demonstrated the capability to perform a variety of tasks without being constructed with specific knowledge of the rules governing the task \cite{Lecun2015,Wetzstein2020}.}
Important ML performance metrics, such as speed and energy efficiency, depend on the computing platform on which the ML algorithm operates.
Accordingly, researchers have been motivated to find platforms and associated algorithms that optimize these metrics.
{In this regard, reservoir computing (RC) \cite{Maass2002,Jaeger2004,Lukosevicius2009,Lu2017}, a type of ML that we describe in Sec. II, has attracted attention because it can be realized in a variety of physical forms \cite{Fernando2003,Appeltant2011,Larger2012,Paquot2012,Soriano2015,Canaday2018,Laporte2018,Tanaka2019,Laporte2020,Rafayelyan2020,Marcucci2020,Chembo2020,Paudel2020,Porte2021}.}

Based on the preceding motivation, we present in Sec. III an implementation of reservoir computing \cite{Jaeger2004,Lukosevicius2009,Torrejon2017,Canaday2018,Nakajima2019,Zhong2021,Gauthier2021} that utilizes the complex response of short wavelength modes in a reverberant cavity as the reservoir.
When the wavelength of the fields in a cavity is much smaller than the size of the cavity, the wave field has effectively a high degree of freedom.
Equivalently, in this `short wavelength' regime the field can be thought of as a superposition of many modes: the number of which is determined by the bandwidth of the time dependent signals to be produced and the spectral mode density of the cavity.
{This number of participating modes characterizes the effective amount of information contained in a specification of the reservoir state at a given time.
As such, it provides an upper limit on the information handling capacity and computational power of the RC.
In practice, however, this upper limit may far exceed what is actually realized.
We shall loosely refer to the realized number as the RC ``size''.
}

{In this paper, we present a proof-of-principle experimental demonstration of a short wavelength wave-based RC system operating in the microwave regime.}
Most importantly, by exploiting the high degree of freedom of the wave fields, and by introducing several new Reservoir Enhancement Techniques (RETs), we show that a measure of the size of the RC system can be made large, thus greatly enhancing the RC computational power.
The potential computing power and versatility of wave-based RC systems are demonstrated and assessed in Sec. V through experimental and numerical tests on several benchmark tasks.

\section{II. Conventional Reservoir Computing}

\begin{figure*}
\centering
\includegraphics[width=0.9\textwidth]{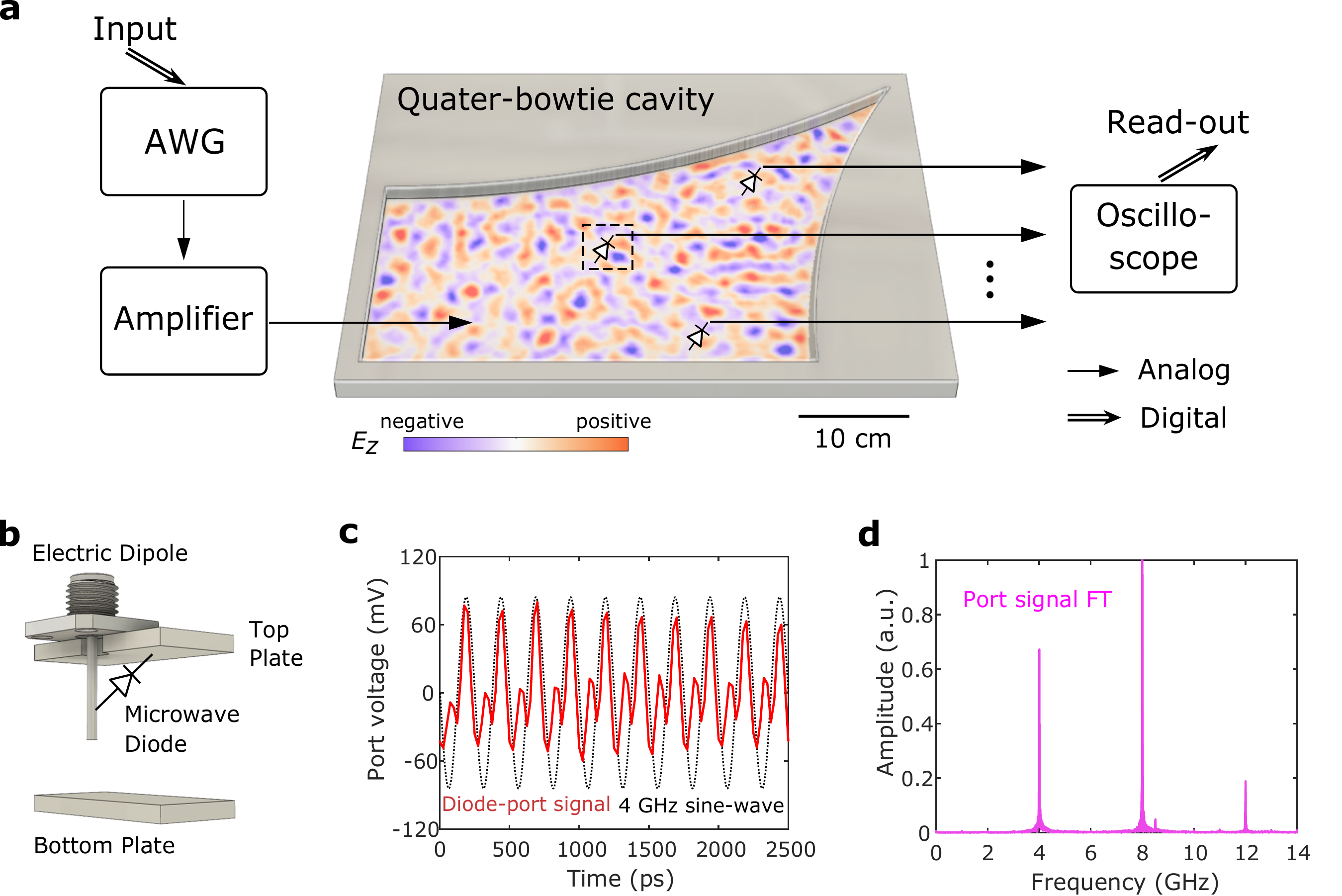}
\caption{\label{fig:ExpSetup} \textbf{Reverberant wave systems for reservoir computing. a}, Schematic of the experimental setup. The input information is transferred from a lab computer to the AWG and injected into the chaotic enclosure through a simple electric dipole antenna. Several diode-loaded antennas are used to probe the EM field, whose voltage signals are measured by the oscilloscope and further transferred to the lab computer and stored.
The cavity shown in panel \textbf{a} is thin in the vertical, $z$-direction, and has a shape in the $(x,y)$ directions {in which the bottom and left walls are straight lines and the upper and right walls are circular arcs.}
This leads to a purely vertical electrical field $E_z(x,y)$ whose complex two-dimensional spatial distribution is shown in panel \textbf{a} via the blue-to-red color coding within the cavity.
\textbf{b}, Schematic of the diode-loaded port. 
\textbf{c}, {The dynamics of diode-port voltage (red) under single frequency (4 GHz) input wave (black) injected at the linear input port.} \textbf{d}, The {Fourier transform (FT)} of the diode-port signal shown in \textbf{c}.}
\end{figure*}

Reservoir computing is a general type of ML whose structure, {in the case of continuous time operation,} can be specified as follows.
{Input variables, in the form of a time $(t)$ dependent vector $u(t)$, drive the evolution of a reservoir state $\hat{r}(t)$.}
The reservoir state $\hat{r}$ is typically a high dimensional vector, and the input $u$ is a much lower dimensional signal vector to which $\hat{r}$ responds.
{(In our case, $\hat{r}$ represents the field within the wave confining structure, e.g., the microwave cavity shown in Fig. \ref{fig:ExpSetup}a.)}

The reservoir state {evolves} according to a reservoir dynamical system $f$,
\begin{equation} \label{eq:eq1}
    \frac{d\hat{r}(t)}{dt} = f(\hat{r}(t),u(t)).
\end{equation}
In Eq. \ref{eq:eq1}, it is assumed that $f$ satisfies the `echo-state' property \cite{Jaeger2004,Lukosevicius2009}, which requires that for any input {time series $u(t)$, $\hat{r}(t)$ becomes independent of the initial condition $\hat{r}(0)$ as $t$ becomes large.}
A time series of output vectors $r(t)$ of dimension $N_r$ is derived from the reservoir state $\hat{r}(t)$ via a function $g$,
\begin{subequations} \label{eq:eq2}
\begin{equation}
    r(t) = g(\hat{r}(t)).
%\tag{\ref{eq:eq2}}
\end{equation}
The overall RC system output, denoted by a vector $s(t)$ is taken to depend linearly on the vector $r(t)$ 
\begin{equation}
    s(t) = W\, r(t)
%      \label{eq:eq2_b}
\end{equation}
\end{subequations}
where $W$ is a rectangular matrix, and the dimension $N_r$ of $r(t)$ is typically large compared with the dimension of the vector $s(t)$.
The elements of the matrix $W$ are viewed as adjustable parameters whose values are chosen through a `training' procedure, whereby, based on training data consisting of examples $(u, s')$ of inputs $u(t)$ and the corresponding desired resulting outputs $s'(t)$,
$W$ is determined by minimizing the deviations of $s(t)$ (the actual RC output) from $s'(t)$.
Heuristically, the basic assumptions of reservoir computing are that, (i) if the dimension $N_r$ is large, and the $N_r$ individual elements of the time-dependent vector $r(t)$ evolve in diverse ways, then the best fit of $s(t) = Wr(t)$ to $s'(t)$ will indeed be a very good approximation to the time varying vector $s'(t)$; and (ii) following training, the RC system outputs will continue to give a good approximation to the outputs $s'(t)$ that would be desired for the post-training inputs.
{Based on item (i) we shall view $N_r$ as quantifying the notion of RC system size.}
Operationally, item (ii) is promoted by the use of training regularization (typically, for RC, ridge regression) meant to prevent over-fitting \cite{Lukosevicius2009}.
In general, either the function $f$, or the function $g$, or both should be nonlinear to allow the RC system to perform a wide variety of nonlinear tasks \cite{Grigoryeva2018}.
We note that, by virtue of the linear relation $s = W\, r$, the training of a RC system reduces to a simple linear regression.
Thus, the training of an RC can be very fast \cite{Vlachas2020}.

\section{III. Reverberant wave RC}

Figure \ref{fig:ExpSetup} shows the proof-of-principle experimental microwave reverberant wave-based RC system considered in this paper.
In Fig. \ref{fig:ExpSetup}a, digital input signals are transformed into analog waveforms by an {Tektronix 70000A AWG (arbitrary waveform generator) and stored for both the training and testing time-series data sets.
{These analog waveforms are exactly the same as the digital input signal stream.
Note that no signal formatting or masking is applied at this step}.
The waveform is then amplified by the RF-Lambda 2-18GHz amplifier (RFLUPA0218G5)}, and injected into a wave chaotic microwave cavity with an area $A=0.115 m^2$ \cite{So1995,Hemmady2005,Zhou2019,Ma2020,Ma2020b}.
{The shape of the cavity is formed from two straight walls intersecting at a right angle and two additional circular arc shaped walls.
The characteristic length of the cavity is of $A^{0.5} \sim 35cm$ and thus the cavity is electrically large compared to the nominal operating wavelength ($\sim 5cm$).
With a height of $d=0.79cm$, the cavity can be considered to be quasi-2D because the electric field is polarized in the z-direction for $f<c/(2d)=18.9GHz$.
Microwave absorbers are employed inside to alter the fading memory of the system.
For tasks running at $\sim GHz$ rates, the cavity decay time is in the $\sim ns$ range.
The port voltage signals are measured with an oscilloscope (Agilent DSO91304A) with a sampling rate of 40Gs/s.}

The time scale of the input {(signals vary on the $\sim 100ps$ scale)} is such that the input stream will excite hundreds of cavity eigenmodes.
The enclosure is designed to act as a two-dimensional `quarter bowtie-shaped' geometry so that the rays (straight line orbits between specular reflection from the cavity walls) show chaotic dynamics {(i.e., ray orbits with nearby initial conditions typically diverge exponentially as they propagate and experience successive reflections from the cavity's enclosing walls).}
The EM field emerges as the real-time superposition of all system modes, sampled at discrete locations (the $N_r$ output ports) in the system.
{To include nonlinearity in to the otherwise linear cavity system, we installed high switching speed diodes (Infineon BAS7004) at the output ports} (Fig. \ref{fig:ExpSetup}b) \cite{Zhou2019}. 
This nonlinearity is demonstrated by the port voltage signal distortion of a sinusoidal wave {injected at the linear input port} (Fig. \ref{fig:ExpSetup}c) and {the resulting} higher harmonic responses (Fig. \ref{fig:ExpSetup}d).
Thus, the complexity of the {reverberant wave} reservoir is ensured by the {short wavelength sensitivity} property {of the cavity fields} and the nonlinear elements {connected at the ports}.

In analogy with the conventional RC, {the reservoir layer} $\hat{r}(t)$ is the field distribution within the cavity {at time $t$}, the function $f$ {describes} the field dynamics in the cavity {and to some extent the action of the diodes at the ports}, $N_r$ is the number of available measurement channels, and $g$ is the nonlinear function realized by the {ports and} diodes \cite{Rafayelyan2020}.
The geometrical simplicity of the cavity is in marked contrast to network implementations of RC where the complexity of the function $f$ is built from a complex network topology.
This simplicity suggests the possibility of fabrication, and mechanical robustness advantages of the wave-chaos approach.

Here the $N_r$ diode-loaded ports serve as the observable reservoir nodes {and hence a measure of the RC size.}
The $N_r$ signals at the ports are recorded on the oscilloscope and transferred to a lab computer for off-line training (Fig. \ref{fig:ExpSetup}a).
The number of wave outputs ($N_r$) of the wave-based RC can, in principle, be increased up to the resolution limit of the operating wavelength.
Challenges, however, arise because the total number of measurement ports ($N_r$) in a physical RC may be limited in practice.
For example in our experiment, the number of output ports is limited by the number of recording channels on the oscilloscope (Fig. \ref{fig:ExpSetup}).

\begin{figure*}
\centering
\includegraphics[width=1\textwidth]{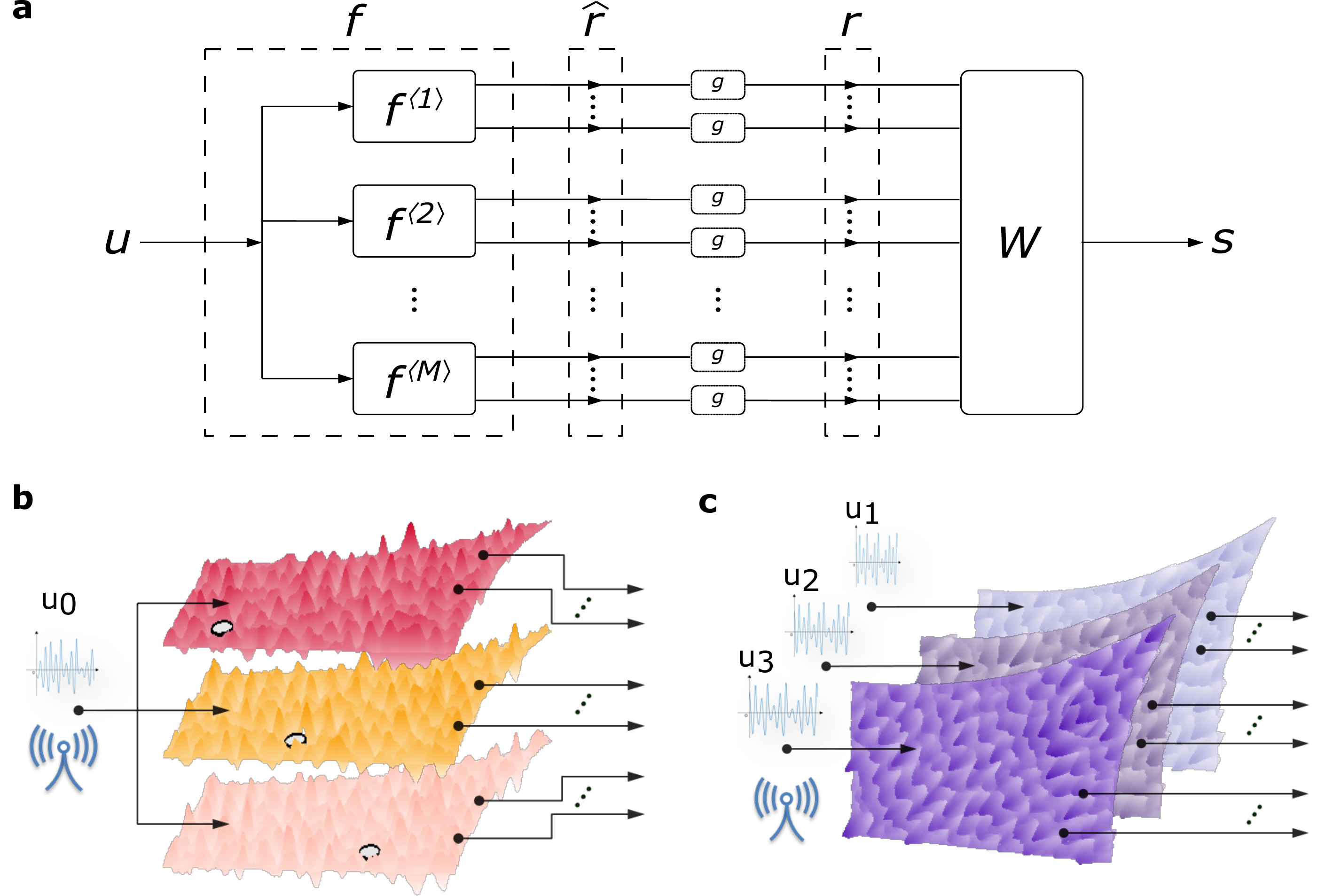}
\caption{\label{fig:virtualnode} \textbf{Reservoir Enhancement Techniques (RETs).}
\textbf{a} The proposed architecture for $f$ and $r$ (Eqs. \ref{eq:eq3} to \ref{eq:eq5}). 
\textbf{b} and \textbf{c} show schematics of the boundary condition method and the frequency-stirring method, respectively. 
The first method may be realized by the translation of a metallic perturber shown as the cylinders. 
Under the same input waveform ${u_0}$, uniquely different evolutions of the EM fields inside the enclosure area are created by means of translating the perturber to new locations.
In \textbf{c}, the frequency stirring technique utilizes small changes of center frequency of a given input waveform to create new wave field configurations. 
In the experiment, the input wavelet is stretched/shrunken by different amounts (forming the inputs, $u_1, u_2,$ and $u_3$) in order to shift the center frequency.
Each output arrow of panels \textbf{b} and \textbf{c} represents one of the $N_r$ components $r^{(i)}$ of the vector $r$.
}
\end{figure*}

\section{IV. Reservoir Enhancement Techniques}

To address the issue of the limited number of output ports {and replicate the RC performance of a cavity with more ports,} we introduce what we call Reservoir Enhancement Techniques (RETs).
To formulate our description of RETs, we consider a specific convenient structure for the $N_r$-dimensional function $f(\hat{r},u)$ in Eq. \ref{eq:eq1}.
In particular, we let $N_r = MN_r'$ and write the $N_r$-dimensional vector $ r = g(\hat{r})$ as the concatenation of $M$ component vectors each of dimension $N_r'$.
Denoting these component vectors $r^{\langle1\rangle}$, $r^{\langle2\rangle}$, ..., $r^{\langle M\rangle}$, we have
\begin{equation} \label{eq:eq3}
    r = \left[r^{\langle1\rangle}; r^{\langle2\rangle}; \cdots; r^{\langle M\rangle} \right],
\end{equation}
Correspondingly, we write $f$ as
\begin{equation} \label{eq:eq4}
    f(\hat{r},u) = \left[ f^{\langle1\rangle}(\hat{r}^{\langle1\rangle},u);  f^{\langle2\rangle}(\hat{r}^{\langle2\rangle},u);  \cdots; f^{\langle M\rangle}(\hat{r}^{\langle M\rangle},u)   \right],
\end{equation}
so that
\begin{equation} \label{eq:eq5}
    \hat{r}_{n+1}^{(i)} = f^{(i)}\left( \hat{r}_{n}^{(i)}, u(t) \right).
\end{equation}
{Thus, each of the $\hat{r}^{(i)}$ evolves independent of the others, except from the mutual dependence on the input stream $u(t)$}.
{Comparing Eq. \ref{eq:eq5} with Eq. \ref{eq:eq1}}, we consider each of the components $\hat{r}^{\langle i\rangle}$ of $\hat{r}$ to be the {wave field distribution in a hypothetical cavity described by $f^{\langle i\rangle}$}.
The idea, illustrated in Fig. \ref{fig:virtualnode}a, is to increase the diversity of the vector $r$ {by increasing its dimension} above that for $M=1$ (ideally by the factor $M$) so that $W r$ can better fit the desired time-dependence of $s$.
Accordingly, we desire the $\hat{r}^{\langle i\rangle}$ to be very different for different $i$, which implies that the $f^{\langle i\rangle}$ must be different for different $i$.

This seems to present a challenge for physical implementation, as it seems to require $M$ different physically constructed cavities.
However, we can achieve Eq. \ref{eq:eq4} with very different $f^{\langle i\rangle}$ and large $M$ by taking advantage of the hallmark property of {short wavelength reverberant waves}, namely, extreme sensitivity of the field distribution within the cavity, and hence also $f^{\langle i\rangle}$, to small changes in the cavity boundaries.
Thus, in our experiments in this paper we will make use of {this sensitivity} through two alternative techniques allowing physical implementation of Eqs. \ref{eq:eq3} and \ref{eq:eq5} using a \textit{single} cavity.
We call these two techniques the `boundary condition method' (Fig. \ref{fig:virtualnode}b) and the `frequency stirring method' (Fig. \ref{fig:virtualnode}c).
As a proof-of-principle demonstration of the `boundary condition method', we place a small conducting circular cylinder in the cavity and move it to different locations, each location corresponding to a different $f^{\langle i\rangle}$.
In the `frequency stirring method' we uniformly scale the duration of the input signal time dependence by a factor $\beta^{\langle i\rangle}$, $t \rightarrow \beta^{\langle i\rangle}\,t$; since the medium within the cavity is air (essentially equivalent to vacuum), the waves are nondispersive, and scaling time by $\beta^{\langle i\rangle}$ is equivalent to scaling distances characterizing the shape of the cavity walls uniformly by the factor $(\beta^{\langle i\rangle})^{-1}$.
{For an input waveform centered at $f_0=4GHz$, we translate the perturber (a conducting cylinder of radius $1.5cm$ and height $0.75cm$) by 1cm ($\sim 0.2\lambda_0$) or shift the center frequency $f_0$ by $\Delta f=100MHz$ ($\sim 3$ resonator eigenmodes) to create a new reservoir.}

Figs. \ref{fig:virtualnode}b and c show snapshots of the complex electric field $E_z$ landscape {within the cavity obtained} from time-domain simulations.
These simulations confirm that small variations of system boundary conditions, as well as small input time stretching or shrinking, results in drastically different wave {field spatial distributions}.
Virtual new unit reservoirs each having distinct temporal dynamics are thus created.
{The application of RET does require a longer operational time to conduct measurements with a single reservoir, and a requirement to store all measured results and combine them later.
However, our RET serves as a unique way to enhance the reservoir size without making new hardware, new ports, or even new cable connections inside the system.}
Although we have not implemented them in the experiments reported in this paper, there are other ways of physically achieving Eqs. \ref{eq:eq3}-\ref{eq:eq5} with a single cavity.
E.g., one attractive possibility is to place a metasurface with electronically programmable surface impedance elements on a portion of the cavity wall \cite{DelHougne2018,Nadell2019,Frazier2020,Qian2020} and electronically switch between many different surface impedance configurations.
This may be regarded as an electronic implementation of our `boundary condition method' which does not utilize the practically problematic process of physically moving a perturbing object within the cavity.
{We note that related techniques for enhancing the reservoir size is also proposed in Refs. \cite{Grigoryeva2016,Freiberger2020,Zhong2021,Pauwels2021}.}

\begin{figure*}
\centering
\includegraphics[width=1\textwidth]{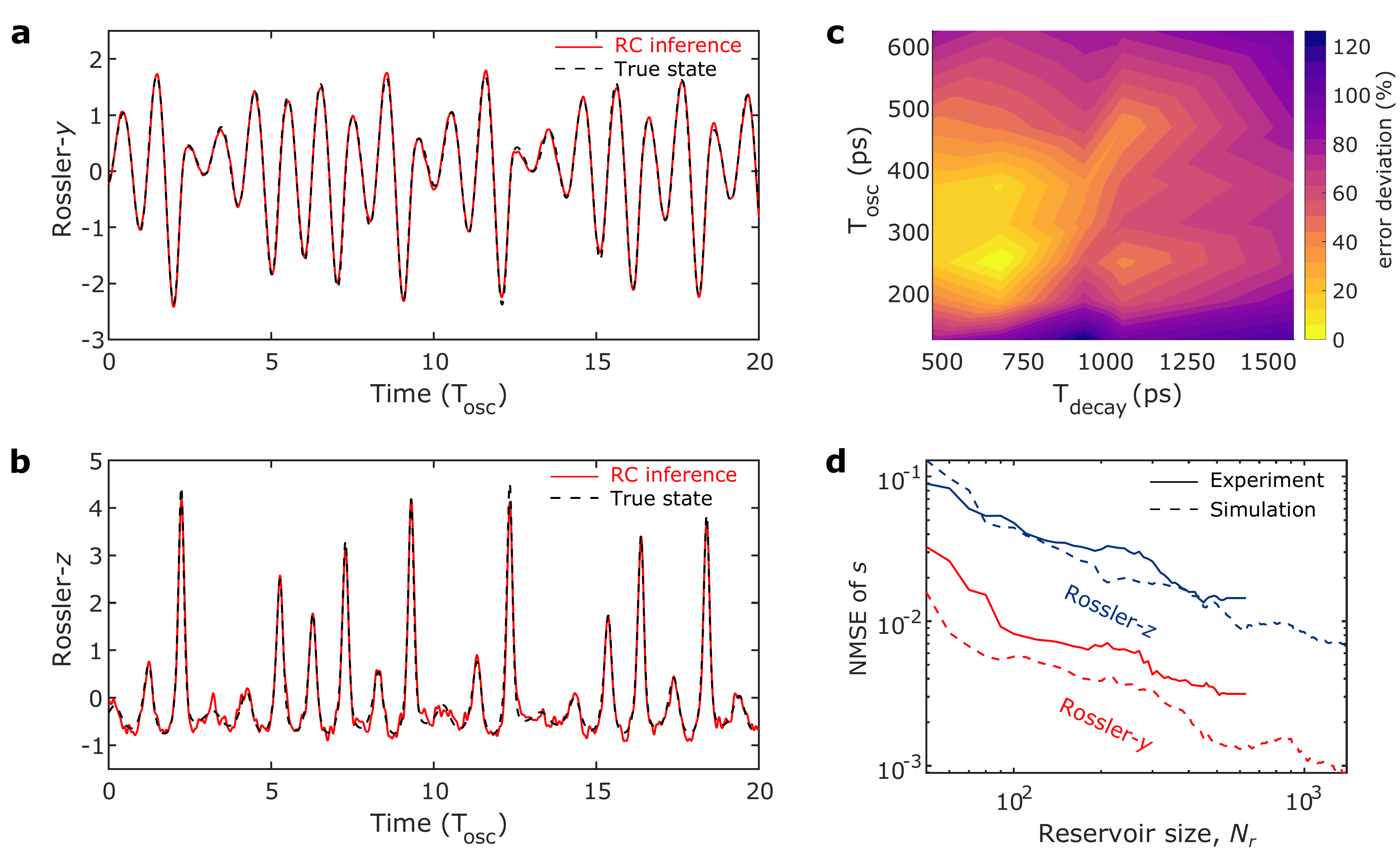}
\caption{\label{fig:Ross}
\textbf{Observer task test results for the R\"{o}ssler system}.
\textbf{a} and \textbf{b}, the RC inferences of $y(t)$ and $z(t)$ (red solid line) versus time in units of $T_{osc}$ agree with their true values (black dashed line).
\textbf{c} {A plot at $N_r = 90$ $(N_0 = 3, N_b = 30)$ of the} percent deviation of the R\"{o}ssler observer task NMSE, $\sum_n|s(t) - s'(t)|^2/\sum_n |s(t)|^2$, from the NMSE with optimal parameters $\left( T_{osc}, T_{decay} \right) = (250, 700)$ as a function of the input duration $T_{osc}$ and the system decay time $T_{decay}$. 
\textbf{d} The normalized mean square error in $s$ versus $N_r$ the dimension of $r$, {which is varied, e.g., for the experimental (solid) curves, by starting at a maximum value of $N_r = 630$ (corresponding to $N_0 = 3, N_b = 30, N_f = 7$), and then randomly removing virtual outputs to successively lower $N_r$.}
}
\end{figure*}

\section{V. Results}

We have experimentally investigated the effectiveness of our {reverberant wave approach} to implementing RC on several different tasks.
Ensembles of new reservoirs are created by translating a metallic perturber and/or changing the {oscillation period of the input signal} from the AWG.
In our experiment, a combined RC of size $N_r$ is given by 
\begin{equation} \label{eq:eq6}
    N_r = N_0 \, N_b \, N_f,
\end{equation}
where $N_0, N_b$, and $N_f$ represent the number of measurement channels, the number of applications of the boundary condition perturbation RET, and the number of applications of the frequency stirring RET, respectively.

\begin{figure*}
\centering
\includegraphics[width=1\textwidth]{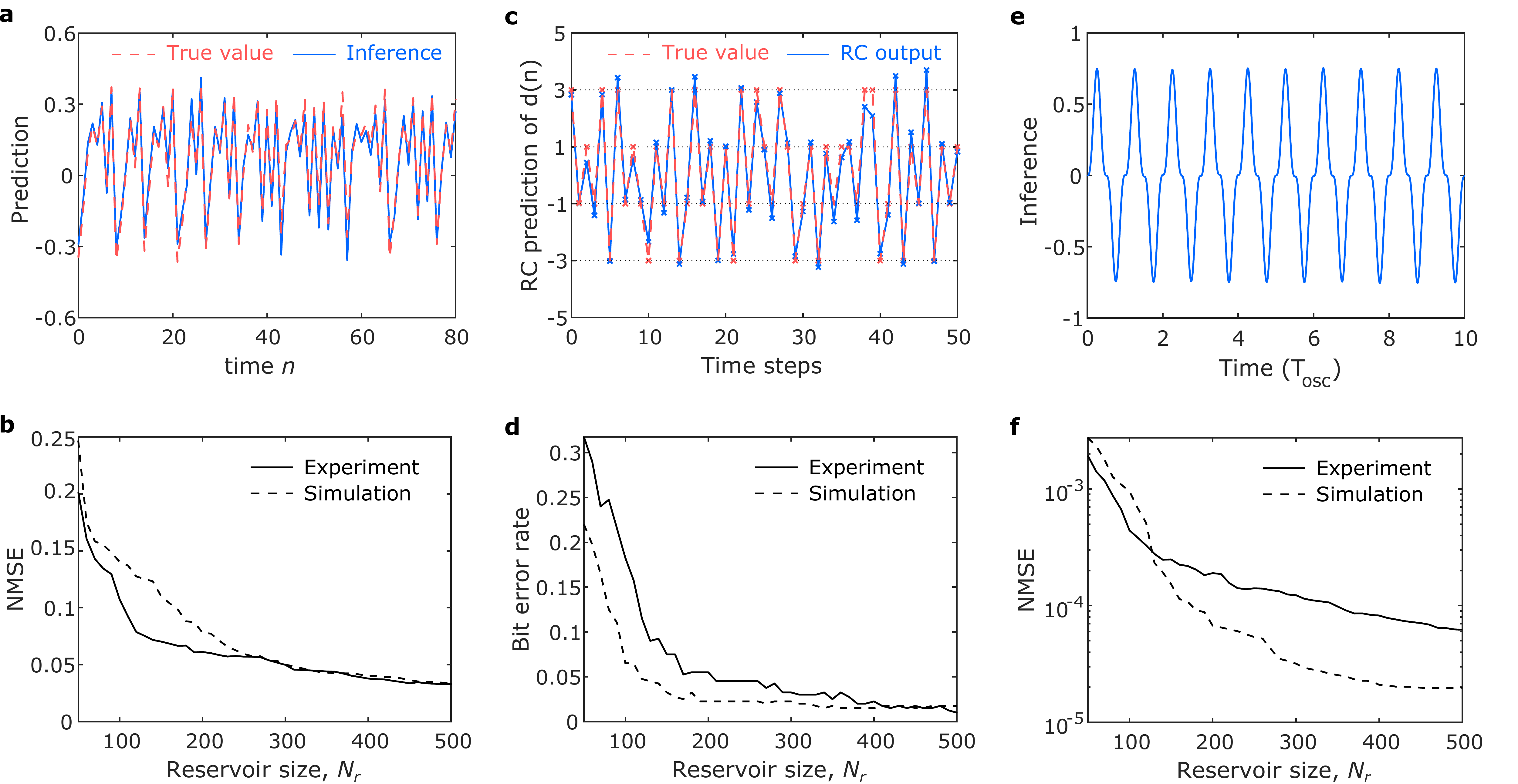}
\caption{\label{fig:tasks}
\textbf{H\'{e}non map observer task performance.} 
\textbf{a} The true orbit $y_n$ (dashed red) and its RC inference (blue) as a function of time step $n$ for a RC size of $N_r = 540$ obtained using 3 cavity ports, 30 implementation of the boundary RET and 6 implementations of the frequency stirring RET ($N_r = 3\times 30 \times 6 = 540$).
\textbf{b} The RC inference normalized error versus the reservoir size, $N_r$.
\textbf{Nonlinear channel equalizer (NCE) task performance.} 
\textbf{c} The RC performances (blue) for the NCE task.
The targets are shown as red dashed lines.
In \textbf{d}, we show the simulated (dashed) and experimental (solid) RC performance as a function of the reservoir size $N_r$.
\textbf{Function simulator task performance.} 
\textbf{e} The RC performance (blue) for the function simulator task.
In \textbf{f}, we show the simulated (dashed) and experimental (solid) RC performance as a function of the dimension of the reservoir.
}
\end{figure*}

For the experiments described below we used the bowtie cavity with $N_0 = 3$ output ports.
Without using RETs we find that, for all of the 5 tasks tested, the RC system fails to give useful results.
However, with RET implemented, the performance of all 5 tasks improves, becoming better as the effective RC system size $N_r$ increases (Figs. \ref{fig:Ross}d, \ref{fig:tasks}b, \ref{fig:tasks}d, and \ref{fig:tasks}f).
For the 5 examples tested we found that using RET to sufficiently boost $N_r$ resulted in excellent performance.
In the rest of this section we give results of our tests.

\textit{Example 1: The observer task applied to the continuous time chaotic R\"{o}ssler attractor.}
For the observer task \cite{Lu2017}, the RC system is expected to infer the time variation of unobserved state variables of a dynamical system (in this example, the $y(t)$ and $z(t)$ components of the chaotic R\"{o}ssler system) based only on observations of a subset of the system {state variables} (in this example the R\"{o}ssler $x(t)$ component).
{The R\"{o}ssler system is governed by these equations,
\begin{equation} \label{eq:Rossler}
\begin{aligned}
    \Dot{x} &= -y-z,\\
    \Dot{y} &= x+ay,\\
    \Dot{z} &= b+z(x-c),\\
    \end{aligned}
\end{equation}
where $a=0.5, b=2.1, c=3.5$, and the over-dot denotes derivative with respect to time.}

Like other ML methods, this prediction task is carried out without knowledge of the equations of motion of the R\"{o}ssler system, using only a finite duration of data of all three components for training. 
The performance of the RC inference $s=(y,z)$ is evaluated using the normalized mean square error, e.g., defined for the $y$-variable as NMSE $= \sum_n[y_n - y_n']^2/\sum_n y_n^2$, where $y'$ and $y$ denote true and RC-inferred values, respectively, and $n$ is the time index of the testing set data, $y_n=y(n \, \Delta t)$ with $\Delta t$ chosen small enough compared to $T_{osc}$ (the time scale for the variation of the R\"{o}ssler state) that $y$ at the discrete times $n \, \Delta t$ traces out a good approximation of the continuous variation of $y(t)$.
The NMSE of the R\"{o}ssler z component is computed analogously.
Here $T_{osc}$ is defined by first noting that the frequency power spectrum of the RC input (here the R\"{o}ssler $x$ series) shows a pronounced well-defined lowest spectral component, and $T_{osc}$ is defined as the period corresponding to the frequency at the peak of this spectral component.
{In the training period, the input is the R\"{o}ssler $x$ component with $\sim 200$ oscillation periods $T_{osc}$, and the testing period includes $\sim 50$ periods.}
We also note that we can vary $T_{osc}$ by application of the previously mentioned time scaling of the RC system input ($t \rightarrow \beta t$).
With RET, a reservoir of $N_r = 630$ output channels is created and this RC and achieves NMSE = (0.003, 0.014) for inference of the R\"{o}ssler y and z components, respectively.
{This combined RC size ($N_r = 630$) is achieved from 30 applications of the RET boundary method ($N_b = 30$) and 7 applications of the RET frequency stirring method ($N_f = 7$), which, from Eq. \ref{eq:eq6}, when combined with the 3 existing ports of the cavity ($N_0 = 3$) yields $N_r = 3\times 30\times 7 = 630$.}
In Figs. \ref{fig:Ross}a and b the inferred R\"{o}ssler $y$ and $z$ components, obtained using the $N_r = 630$ RC, are plotted as red solid lines and accurately reproduce their corresponding true values (black dashed lines).

Fig. \ref{fig:Ross}c shows a heat-map of the reservoir performance, with $N_r = 90$, in the inference of $s=(y,z)$ versus the reservoir parameters $T_{osc}$ and $T_{decay}$.
Here $T_{decay}$ denotes the exponential damping time of waves of frequency $2\pi/T_{osc}$ in the {undriven} cavity, and is varied by placing dissipative material within the cavity.
{For an empty cavity, the measured decay time is $\sim2.14ns$.}
Thus the shortest decay time (left boundary of the heat map) corresponds to the case of the most lossy cavity.
As shown in Fig. \ref{fig:Ross}d, as $N_r$ increases {via applications of RETs (see figure caption)}, the RC performance is greatly improved.
%{In Sup. Mat. Section I, we studied the effect of different training set length on the testing set performance.}

\textit{Example 2: The observer task for the discrete-time chaotic H\'{e}non-map.}
Results for the RC observer task applied to the two-dimensional discrete-time chaotic H\'{e}non-map, $(x_{n+1}, y_{n+1}) = (1-1.4x_n^2+y_n, 0.3x_n)$, are shown in Fig. \ref{fig:tasks}a and b, where the variable $x$ is observed and $y$ is inferred.
Each input value of the discrete $x$ series is sampled for $T_{bin} \sim 60ps$.
%{More explicitly, each input value $x$ is represented by a bin-shaped waveform for a duration of $T_{bin}$.}
The decay time of the system is fixed at $T_{decay} \sim 600ps$.
The training waveform of the H\'{e}non-map has a length of 4000 time steps, and the testing waveform is set to 1000 bins.
{We see that the RC-inferred values of $y$ (plotted in blue) agree well with the true values (plotted as a dashed red curve). 
For these results, RETs yielding $N_r=540$ (see figure caption) were employed.}

\textit{Example 3: The nonlinear channel equalization task.}
For the nonlinear channel equalization (NCE) task, the goal is to recover a random 4-level symbol sequence from a noisy sequence which simulates the received signal sent through a nonlinear multi-path RF channel.
For the 4-level symbol sequence, a random series $d(n)$ is chosen between the levels $\{-3,-1,1,3\}$, and the RF channel signal is assumed to be $q(n) = 0.08d(n+2) - 0.12d(n+1) + d(n) + 0.18d(n-1) - 0.1d(n-2) + 0.091d(n-3) - 0.05d(n-4) + 0.04d(n-5) + 0.03d(n-6) + 0.01d(n-7)$.
The RC system input for this NCE task is the channel signal $q(n)$, and the task is to retrieve the original four-level random $d(n)$ series.
In the experiment, an input speed of $T_{bin} \sim 60ps$ is adopted, and the training/testing set includes 4000/1000 time steps, respectively.
{The decay time of the system is fixed at $T_{decay} \sim 600ps$.}
The direct RC output is regularized to the nearest level.
Results for this test are shown in Fig. \ref{fig:tasks}c and d.

\textit{Example 4: The function simulator task.}
For the function simulator task, the RC is expected to output any periodic waveform that is desired.
For this purpose, we take the input to be a sinusoidal waveform with the period of the desired waveform.
In our test example we take the desired waveform to be the cube of the sine wave, and we train the RC system to give this output.
{The decay time of the system is fixed at $T_{decay} \sim 600ps$.}
We employ a $4GHz$ sine wave input with a duration of $\sim 300$ oscillation periods.
The length of the training and testing sets are set to a 80-20 ratio.
Results for this test show very good agreement between the RC output waveform and the target (Fig. \ref{fig:tasks}e and f).
(We have also confirmed that the wave-based RC is able to generate other types of input functions, including two-tone and three-tone signals.)

\begin{figure*}
\centering
\includegraphics[width=1\textwidth]{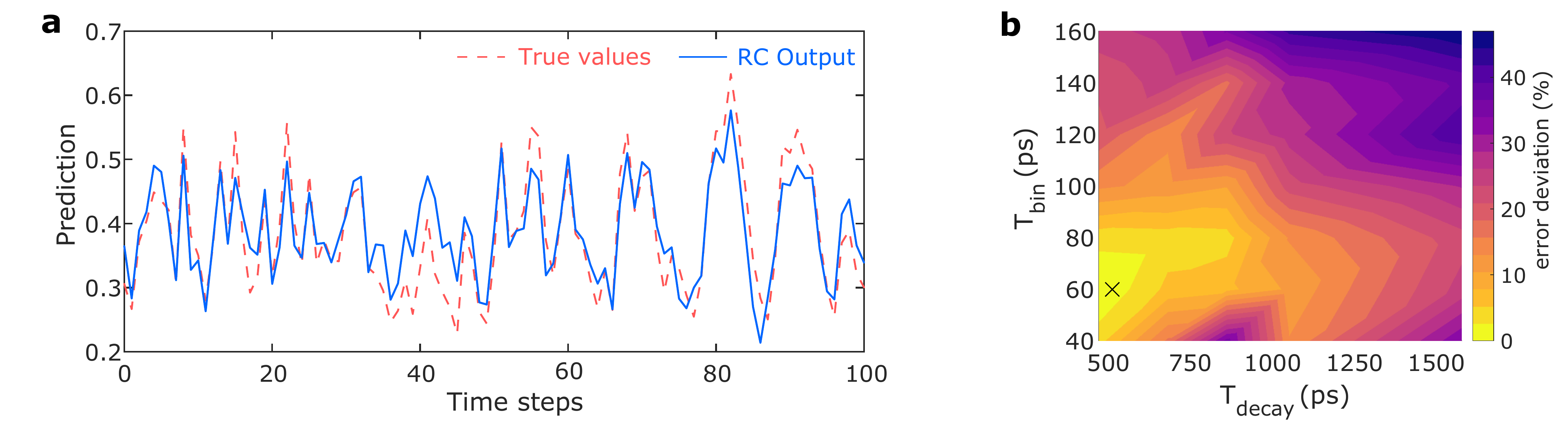}
\caption{\label{fig:narma}
\textbf{a,} NARMA-10 task testing set performance at the optimal system parameters from a {RC size of $N_r = 90$ (obtained via {the 3 cavity ports with} 30 applications of the boundary RET)}, shown as the cross in \textbf{b}.
\textbf{c} shows the deviation of the NMSE for a $N_r = 90$ $(N_0 = 3, N_b = 30)$ RC. The percent deviation is computed with respect to the NMSE of the optimal parameters $\left( T_{bin}, T_{decay} \right) = (60, 550)$ as a function of the input value duration $T_{bin}$ and the system decay time $T_{decay}$}
\end{figure*}

\textit{Example 5: The NARMA-10 task.}
For the 10-time-step nonlinear autoregressive moving average (NARMA-10) task, the input stream $u(n)$ is a random series drawn from the interval $[0, 0.5]$.
{The target output is computed from the following 10th-order nonlinear relationship: $y(n+1) = 0.3y(n) + 0.05y(n)\left[ \sum_{i=0}^9 y(n-i) \right] + 1.5u(n-9)\cdot u(n) + 0.1$. 
Its complex behavior and a 10-state memory requirement make the NARMA-10 task a popular benchmark test for both software and hardware RC \cite{Tanaka2019,Chembo2020}.
The training waveform has a length of 4000 random values, and the testing waveform is set to 1000 values.}
With optimal system parameters (cross in Fig. \ref{fig:narma}b), a RC with $N_r = 90$ achieves a performance of NMSE = 0.034 which compares favorably with that reported in several recent photonic hardware RC implementations (e.g. see \cite{Chembo2020}).
The optimized performance island occurs when $T_{decay} \sim 9T_{bin}$ (Fig. \ref{fig:narma}b).
This empirical observation agrees nicely with the setup of the NARMA-10 task where each output time step is determined by its 10 previous inputs, thus demonstrating the complex memory capacity of the wave-based RC.
{In the experiment, the input is a time-domain waveform where each value is sampled for $T_{bin} \sim 60ps$.
We note that such choices of input waveform speeds are limited by the sampling speed of both the AWG and the oscilloscope.}

Besides experimental tests, we have also conducted electromagnetic (EM) numerical simulations {in CST Microwave Studio} of the physical RC with the same shape cavity, identical degree of system loss, and realistic model of the diodes. 
{The high-dimensional combined RC is realized with the RETs, where the boundary condition perturbation is realized using the cylindrical metallic perturber with the same dimension as the experimental one.}
As shown in Figs. \ref{fig:Ross}d, \ref{fig:tasks}b, \ref{fig:tasks}d and \ref{fig:tasks}f, all of our experimental cases are faithfully simulated with EM simulation numerical tools.
The accurate simulation capability of wave-based RC greatly benefits future RC optimization and follow-up studies.

{We also point out that the ultimate realization of the proposed RC will have as many ports as needed to achieve a desired performance.
Using full-wave simulations, we tested a version of the RC which has $N_r >> 3$ ports installed in one single cavity and found good performance (not shown in the paper).
In experiments, RET allows us to validate our wave-based RC concept under our limited measurement capabilities. 
It would otherwise be impossible for us to demonstrate our hardware RC concept.}
{The RET concepts show that a reservoir consisting of a number of wave-based RCs running in parallel will also work. 
Similar combined RC configurations are also introduced in } Refs. \cite{Nakajima2019, Porte2021, Zhong2021}.

\section{VI. Conclusion}

Benefiting from {the basic nature of short wavelength reverberant wave systems}, our RC scheme shows advantages in its simplified physical structure and insensitivity to structural details. 
The computational performance of the wave based RC, quantified by the testing set error for various benchmark tests, is greatly improved by the expansion of the reservoir size.
We note that the efficacy of the output coupling matrix $W$ may degrade as the RC scattering properties change over time (e.g., due to aging at very long time) \cite{Taddese2010,Taddese2013}.
However, this performance drift can be quickly re-calibrated because the training of the RC is fast.

In summary, we have experimentally demonstrated a physical platform for reservoir computing utilizing the complex dynamics of waves, and found good agreement between experiments and simulations.
By exploiting the fundamental property of the short wavelength systems (i.e., extreme sensitivity of the wave field distribution to perturbations), we formulate techniques for expanding the size and computational power of wave-based RC.
We further demonstrate the effectiveness of our approach by the successful execution of different benchmark tests.
Our general scheme for enhancing the computational power of RC (Fig. \ref{fig:virtualnode}a) may be of general use beyond application to wave-based RC.

\section{Acknowledgements}

This work was supported by ONR under Grant No. N000141912481, AFOSR COE Grant FA9550-15-1-0171, and the Maryland Quantum Materials Center.

\end{spacing}

\end{document}